\documentclass[11pt,letterpaper]{article}
\date{}
\usepackage{subfigure}
\usepackage{subcaption}
\usepackage[authoryear]{natbib}
\usepackage[letterpaper, total={6.5in, 9.5in}]{geometry}
\usepackage{xspace}

\newcommand{\OURS}{CosmoFlow\xspace}

\usepackage{hyperref}
\usepackage{url}
\usepackage{booktabs}
\usepackage{tabularx}
\usepackage{graphicx}
\usepackage{amsmath}
\usepackage{amssymb}
\usepackage{mathrsfs}
\usepackage[scr=rsfso]{mathalpha}
\usepackage{xcolor}

\title{CosmoFlow: Scale-Aware Representation Learning for Cosmology with Flow Matching }

\author{Sidharth Kannan$^1$ ,~~Tian Qiu$^1$ ,~~Carolina Cuesta-Lazaro$^{2,3,4}$,~~Haewon Jeong$^1$\\
\normalsize
$^1$ \textit{University of California, Santa Barbara}\\
\normalsize
$^2$ \textit{The NSF AI Institute for Artificial Intelligence and Fundamental Interactions}\\
\normalsize
$^3$ \textit{Department of Physics, Massachusetts Institute of Technology} \\
\normalsize
$^4$\textit{Center for Astrophysics |Harvard \& Smithsonian} \\
\texttt{\{skannan, tian\_qiu,haewon\}@ucsb.edu, cuestalz@mit.edu}
}

\begin{document}
\maketitle

\begin{abstract}
    Generative machine learning models have been demonstrated to be able to learn low dimensional representations of data that preserve information required for downstream tasks. In this work, we demonstrate that \textit{flow matching}-based  generative models can learn compact, semantically rich latent representations of field level cold dark matter (CDM) simulation data without supervision.
Our model, \textbf{CosmoFlow}, learns representations 32x smaller than the raw field data, usable for field level reconstruction, synthetic data generation, and parameter inference. Our model also learns \textit{interpretable} representations, in which different latent channels correspond to features at different cosmological scales. 
\end{abstract}

\section{Introduction}
% bullet points
% \begin{itemize}
%     \item P1: Representation learning is important for cosmology simulation data because it enables efficient analysis (mention tasks), and a meaningful latent space allows fast generation of data. This poses a dual objective for the model: semantically meaning latent, and high fidelity reconstruction.  
%     \item P2: Existing solutions with fail, since they do not satisfy this dual purpose. The widely used VAE-based methods fail to capture high frequency / small scale details
%     \item P3: The need for better representation learning, and the focus of our paper. 
%     \item P4: Diffusion and FM success so far in image generation world, why this is a promising method for cosmology data representation learning. 
%     \item P5: We propose a FM based representation learning framework. Inspired by physical principles, we intend to design latents that are aware of spatial scale. Inspired the DiTi (cite) solution, we apply latents masking in training, and demonstrate frequency separation ability. 
%     \item bullet points of contribution.
% \end{itemize}

The large-scale structure of the Universe provides one of the most stringent tests of gravity on cosmological scales. Over the past decades, the $\Lambda$CDM cosmological model has emerged as the standard framework for understanding our cosmos, where $\Lambda$ represents the cosmological constant (associated with dark energy) and CDM denotes cold dark matter—which together comprise approximately $95\%$ of the Universe's energy budget. Theoretical predictions of $\Lambda$CDM can now be implemented with remarkable precision in numerical simulations, which capture the formation of the cosmic web: an intricate network where galaxies reside in dense clusters, connected by filamentary structures and separated by vast cosmic voids.

This success, however, presents cosmology with a new challenge. High-resolution simulations like AbacusSummit generate datasets exceeding $2000$ TB, severely constraining our ability to scale training datasets for machine learning applications. Moreover, extracting meaningful insights from these high-dimensional datasets requires models that can effectively navigate the curse of dimensionality.

% visual compare fig - page 1 top right
\begin{figure}[ht!]
    \centering
    \includegraphics[width=0.5\linewidth]{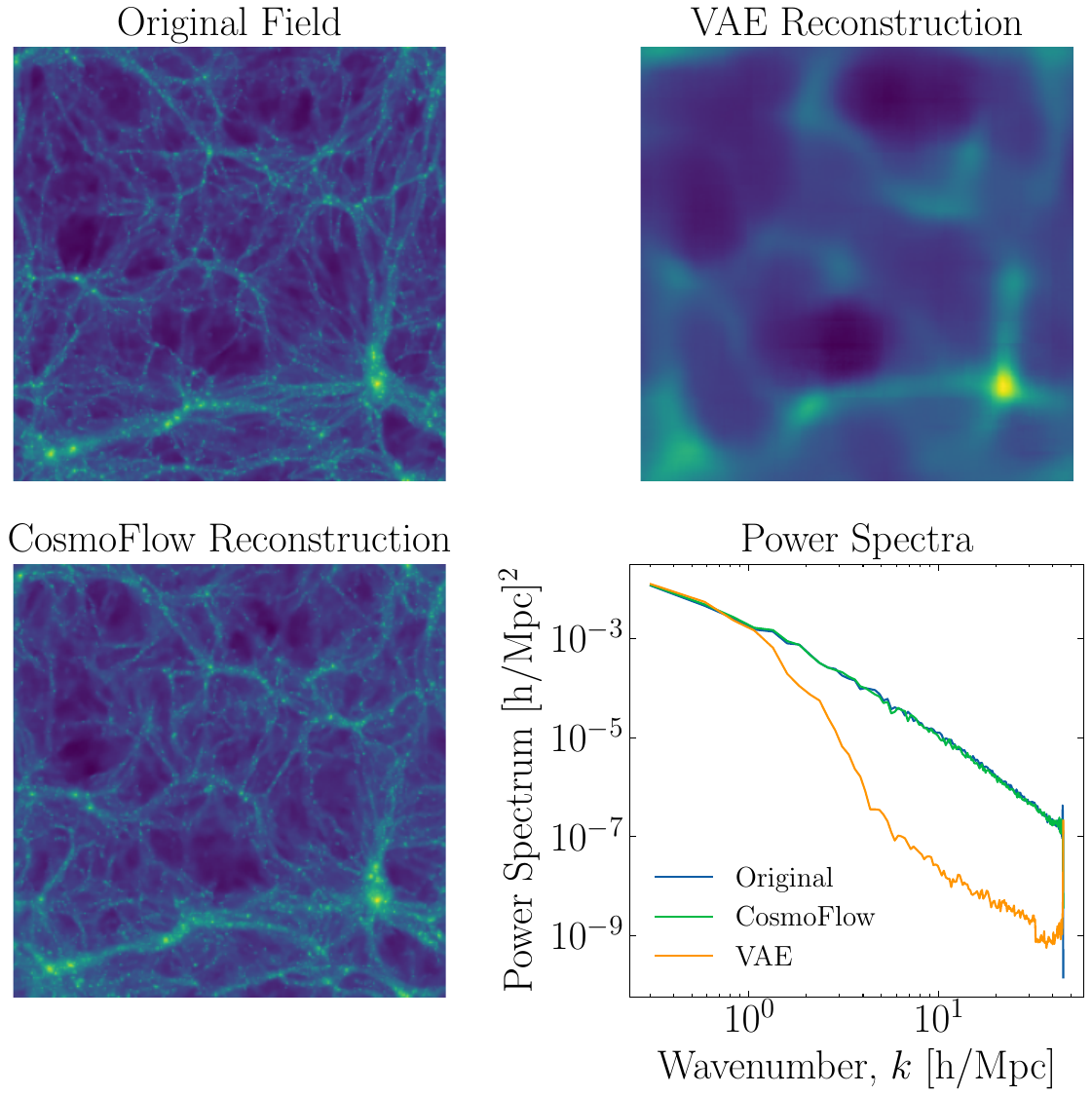}
    % \vspace{-2em}
    \caption{We compare reconstruction quality of our model, \OURS, to the reconstructions produced by a VAE with the same size latent code. While standard VAEs produce blurry reconstructions, our model is able to capture fine detail. We note that the model still deviates from the ground truth at high frequencies.}
    \label{fig:vae-comparison}
     \vspace{-1em}
\end{figure}

Representation learning offers a promising solution by mapping high-dimensional simulation data into low-dimensional representations suitable for downstream tasks such as cosmological parameter inference, anomaly detection, i.e., looking for deviations to $\Lambda$CDM, and data compression. However, the application of representation learning to cosmology faces a fundamental tension between two competing objectives: faithful data reconstruction and semantic information extraction. 

Traditional compression prioritizes minimizing information loss for perfect reconstruction, but many cosmological analyses do not require pixel-level fidelity. Instead, the goal is to capture scientifically relevant information—analogous to how the power spectrum discards spatial phase information while preserving the statistical properties crucial for parameter inference.

Existing approaches to cosmological representation learning fall into two broad categories. Contrastive approaches learn representations by distinguishing between positive and negative example pairs, requiring explicit definitions of which examples should be proximate or distant in latent space. For instance, [\cite{Akhmetzhanova_2023}] defines positive pairs as simulations sharing identical cosmological parameters but differing in initial conditions, while negative pairs correspond to simulations with different cosmological parameters. 

Generative approaches, conversely, learn representations by reconstructing the original data distribution from the latent space. While Variational Autoencoders (VAEs) ~[\citep{kingma2022autoencodingvariationalbayes}] have been widely adopted for this purpose and achieve strong downstream task performance, they typically produce blurry reconstructions that lose critical high-frequency details. Unlike natural images where high frequency details can be perceptually insignificant, a significant amount of cosmological information is present in the small scale structure. 

We adopt a generative framework for the following reasons. First, generative methods eliminate the need for assumptions about what constitutes positive and negative pairs, whereas in constrastive approaches the choice of pairing strategy can substantially impact learned representations. Second, generative models serve multiple purposes: they enable data compression through dimensionality reduction, facilitate fast generation of new, synthetic data, and produce semantically rich representations suitable for downstream cosmological analyses.

Recently, the flow matching paradigm~[\cite{lipman2023flowmatchinggenerativemodeling, lipman2024flowmatchingguidecode, albergo2023stochasticinterpolantsunifyingframework}] has been demonstrated to achieve state-of-the-art performance for generation across image, audio, and other domains. A flow matching model learns to map a noise sample to a data sample, via a time-dependent vector field. In this paper, we apply flow matching to the representation learning problem for cosmology, and demonstrate that it enables the learning of useful and interpretable latent representations.

% In this paper, we pursue self-supervised learning in cosmology with two complementary yet distinct objectives: efficient simulation data compression and the extraction of semantically rich summary statistics for downstream applications. While these goals are related, they represent different philosophies of information preservation. 

% Traditional compression seeks to minimize information loss for faithful data reconstruction—a necessity given that high-resolution N-body simulations like AbacusSummit require X of storage, severely constraining our ability to scale training datasets for machine learning models. However, for many cosmological analyses, perfect reconstruction is neither necessary nor optimal. Instead, we seek representations that capture the most scientifically relevant information—akin to how the power spectrum discards spatial phase information while preserving the statistical properties essential for parameter inference. This approach allows us to learn compressed representations that may sacrifice pixel-level fidelity but retain the semantic structure needed for downstream tasks such as cosmological parameter inference, finding anomalies that deviate from $\Lambda$CDM, and efficient transfer from inexpensive simulations to more computationally demanding scenarios.

We summarize our contributions as follows:
\begin{itemize}
    \item We present \OURS, the first cosmological representation learning model usable for both high quality reconstruction and downstream tasks. We show that our model is able to compress $256\times 256$ pixel field data down to an 8 element vector that can be used to estimate the cosmological parameters with equivalent accuracy as estimation on the raw field data.
    \item We show that our model can be used to generate reconstructions of field data from a latent 32x smaller than the original images, and to generate new, synthetic data for parameter values not in the dataset.
    \item We demonstrate that the inductive biases of flow matching can be used to build a latent space where different parts of the representation correspond to features at different cosmological scales.
\end{itemize}
\begin{figure}[ht!]
  \centering
  \includegraphics[width=0.6
  \columnwidth]{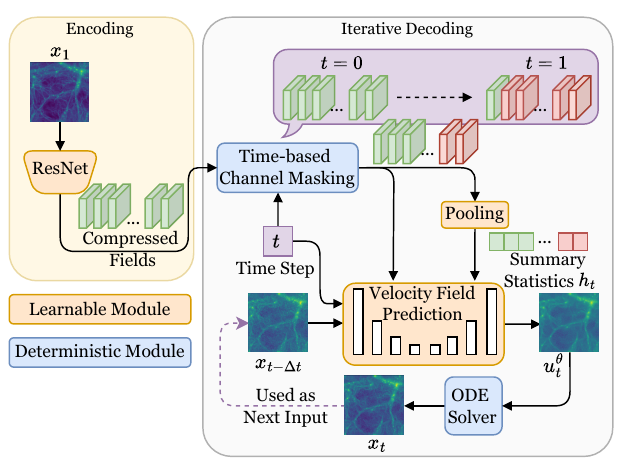}
  % \vspace{-2em}
  \caption{
  \textbf{An overview of \OURS}. A ResNet encodes the input image to compressed fields. During each time step of iterative decoding, the compressed field is masked and passed through a global pooling layer to generate a compact summary statistics vector. Both the masked compressed field and the summary statistics are used as conditions for the UNet-based velocity field prediction. See more details in Sec.~\ref{sec:method}.}
  \label{fig:architecture}
   \vspace{-1em}
\end{figure}
\section{Background and Related Work}
\paragraph{Flow Matching}
% architecture - page 2 top 

Flow matching [\cite{lipman2023flowmatchinggenerativemodeling}] is a framework for generative modelling, closely related to continuous normalizing flows [\cite{chen2018neural}] and the diffusion family of models [\cite{song2021scorebasedgenerativemodelingstochastic, song2020denoising}]. In flow matching, data is mapped from a prior distribution to the data distribution via a \textit{probability flow}, defined via a time-dependent vector field, called the \textit{velocity field}. Samples are then generated by solving the probability flow ordinary differential equation (ODE), with an initial condition $X_0$, randomly sampled from the prior:
\begin{equation}
    \frac{d}{dt}X_t = u_t^\theta(X_t).
\end{equation}
While in general, flow matching allows the prior distribution to take any form, in this work, we use the standard Gaussian, $X_0\sim\mathcal{N}(0, \textbf{I})$. During training, the model attempts to learn a velocity field that maps a sample from the prior to the training sample. This gives rise to the conditional flow matching loss,
\vspace{-0.2em}
\begin{equation}
    \mathcal{L}_\text{CFM} = \mathbb{E}_{t, q(X_1), p_t(X|X_1)}||v_t(x) - u_t(X|X_1)||^2_2.
\end{equation}
In our case, we choose the mapping to be a straight line path. Then, the flow matching loss can be written as a regression loss, between the model output and the difference between the training sample and initial condition:
\begin{equation}
    \mathcal{L} = \frac{1}{N}\sum_n^N ||u^\theta_t(X_t, t) - (X_1-X_0)||^2.
\end{equation}
\begin{figure*}[t!]
  \centering
  \includegraphics[width=0.98\textwidth]{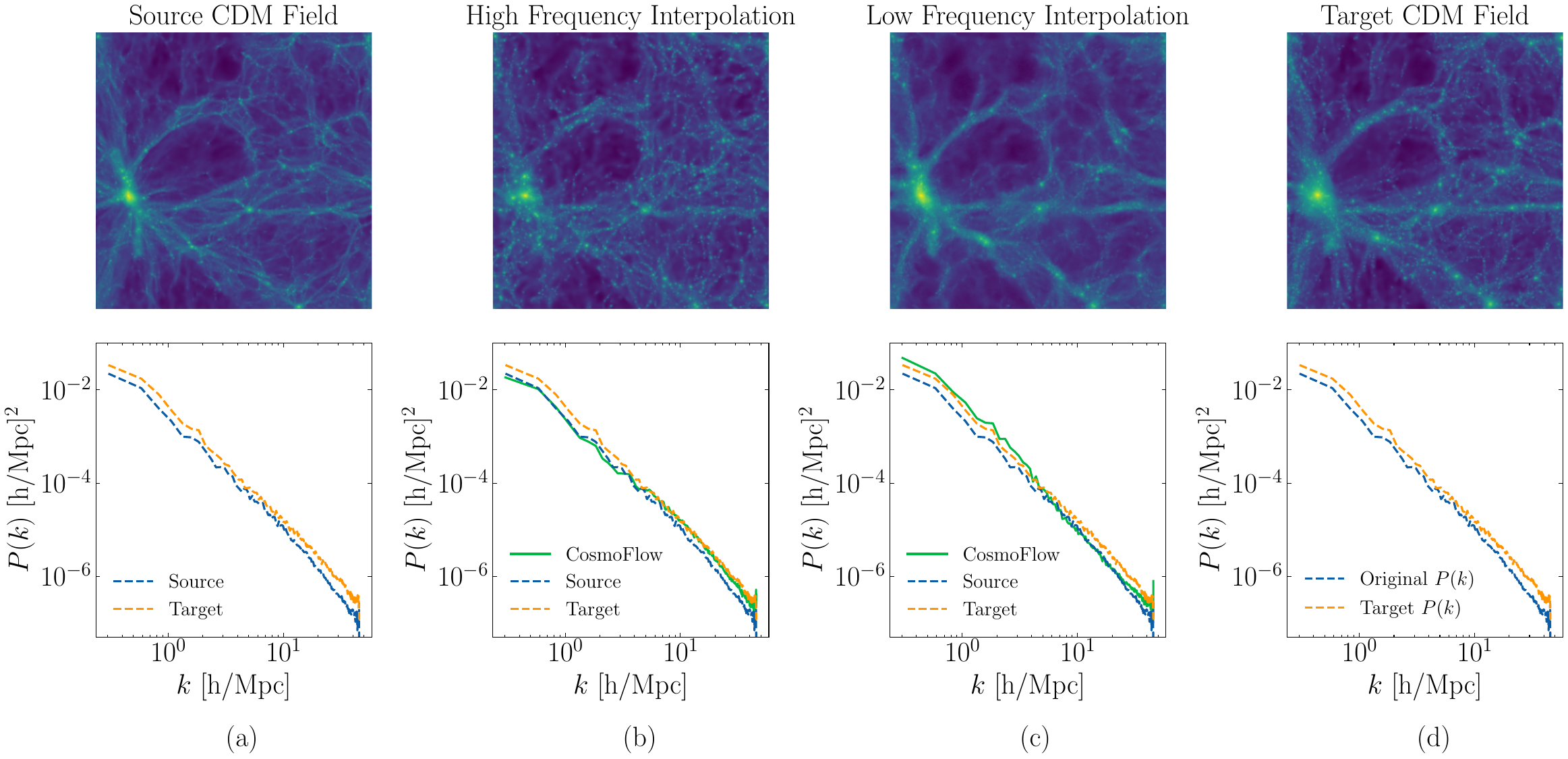}
  % \vspace{-1em}
  \caption{%
    (a) The initial CDM field (b) We interpolate the channels corresponding to high frequencies. The generated image contains many more granular features. Only the high frequency components in the power spectrum rise. (c) We interpolate the channels corresponding to low frequencies. The image smooths, and only the low frequency components in the power spectrum rise.(d) The target CDM field data.%
  }
  % \vspace{-1em}
  \label{fig:interpolation}
\end{figure*}
\vspace{-1.8em}
\paragraph{The CAMELS Dataset}
The CAMELS Multifield Dataset ~[\cite{villaescusa2021multifield}] is a dataset comprised of thousands of hydrodynamic simulations across a wide range of cosmological and astrophysical parameters. The primary goal of  CAMELS  is to elucidate the connection between the various cosmological and astrophysical parameters, and observable features of the universe by enabling the training of machine learning models.

CAMELS contains multiple different simulation suites, each with a different implementation of small scale physics. Each simulation suite captures a range of cosmological fields, including the dark matter distribution, gas distribution, temperature, etc. We train our model using cold dark matter maps from the Astrid hydrodynamic simulation suite [\cite{ni2023camelsprojectexpandinggalaxy}].

\vspace{-0.5em}
\paragraph{Representation Learning in Cosmology}
Recent work in cosmological representation learning has explored both generative and contrastive paradigms. \citet{andrianomena2023latentspacerepresentationscosmological} developed VAEs for cosmological fields, showing strong performance in parameter inference but suffering from the characteristic VAE limitation of blurry reconstructions. \citet{Akhmetzhanova_2023} focused on contrastive approaches, defining positive pairs as simulations with identical cosmologies but different initial conditions, and negative pairs as simulations with different cosmologies.

% We introduce a flow matching-based generative model that addresses the limitations of the above mentioned models while incorporating scale separation as an inductive bias in the latent space—natural given the hierarchical nature of cosmic structure formation. %We evaluate our representations on cosmological parameter inference, anomaly detection (comparing cold and warm dark matter simulations), and data compression, with particular focus on preserving the high-frequency features smoothed out by VAEs.

% \section{Problem Definition}
% The key point of representation learning is to be able to use the compact, learned representations for meaningful tasks. We focus on the following tasks:
% \begin{itemize}
%     \item Reconstruction: We aim to be able to reconstruct the original simulation data with high fidelity, using only the latent.
%     \item Parameter Inference: The latent should encode the cosmological parameters, $\Omega_m, \sigma_8$, which are most strongly represented in the CDM maps.
%     \item Anomaly Detection: The latent representations for anomalous fields should be distinguishable from representations of CDM fields.
%     \item Interpolation: Our latent representations should enable fast generation of realistic field level data with cosmological parameters not represented in the dataset.
% \end{itemize}

\section{Method} \label{sec:method}
\paragraph{Model Architecture}

\OURS is composed of two parts: a) a ResNet [\cite{he2016deep}] based encoder, which produces a lower dimensional representation of the input, used to condition the decoder, and b) a UNet [\cite{ronneberger2015u}] based decoder, which attempts to reconstruct the input image through velocity field estimation. The details of the architecture are shown in Fig. \ref{fig:architecture}.

\vspace{-0.5em}
\paragraph{Learning Spatially Meaningful Latents with Progressive Masking}\label{sec:masking}

Our goal is to design the latent space such that different channels in the compressed field correspond to different scales in the reconstructed image. We do this by employing a version of the framework from \citet{yue2024exploring}. A channel-wise mask is applied to the compressed field such that at $t=0$ (start of generation), all channels are unmasked. As $t$ increases, latent channels are progressively masked out, until at $t=1$, only one remains. This approach is inspired by the inference process in the flow matching family of models (see Appendix \ref{sec:reconstruction-process}), in which a noise sample is iteratively denoised. Low frequency, large scale structure is reconstructed first, while the smaller scale features are refined closer to $t=1$. Thus, the latent channel that remains at $t=1$ encodes the latent information corresponding to the highest frequency.

\section{Results}

\paragraph{Reconstruction}

We demonstrate that our model is able to produce significantly higher fidelity reconstructions than other standard generative models. In particular, we compare against a VAE with the same size latent. We show that CosmoFlow achieves significantly more realistic reconstructions, which is also reflected in the power spectra plot---VAE loses high-frequency information while CosmoFlow preserves all frequencies (see Fig.~\ref{fig:vae-comparison}). We note, however, that our models reconstructions are still lossy; while they are realistic, there are still deviations at high frequency, which can be seen by comparing filaments between the original and reconstruction.

\vspace{-0.6em}
\paragraph{Parameter Inference}
\begin{figure}[t!] % Top of page 3
    \centering
    \includegraphics[width=0.5\linewidth]{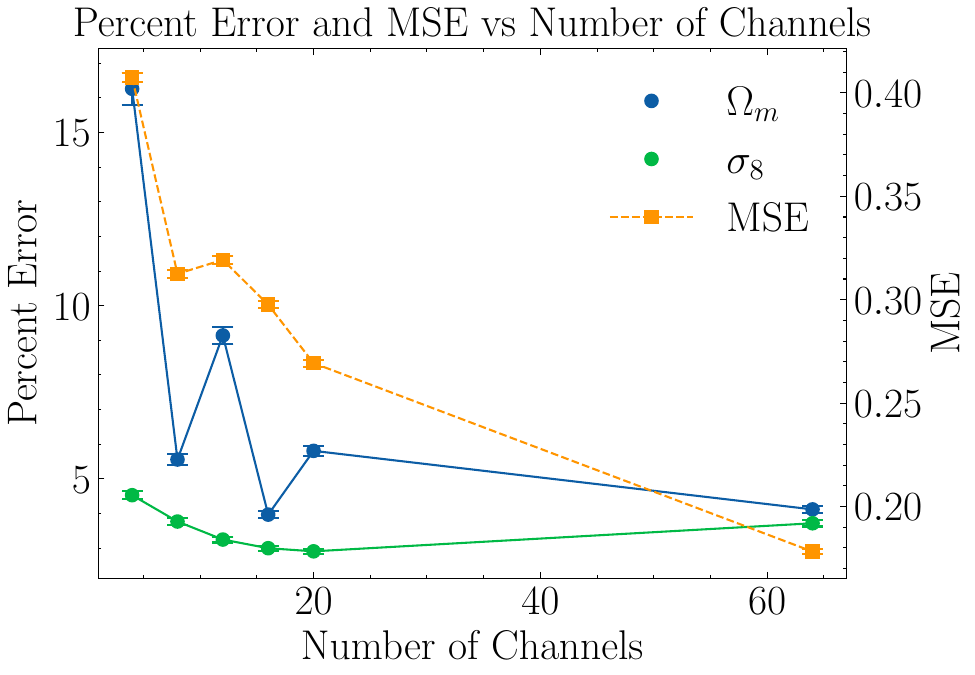}
    % \vspace{-0.7em}
    \caption{We show the effect of varying the number of channels in the compressed field on parameter inference and reconstruction quality. Error bars show the standard error across the validation set. }
    \label{fig:param-est-mse}
    % \vspace{-1em}
\end{figure}
One of the primary applications of representation learning models to cosmology is to learn low dimensional representations which still allow estimation of the posterior distribution of cosmological parameters. We demonstrate that the \textit{summary statistics}, produced by the encoder can be used for parameter inference with similar accuracy to inference on the raw field data. In this work, we focus on predicting the posterior mean of the cosmological parameters instead of the full distribution.

As a baseline, we train a ResNet-18 for parameter inference using the raw field data. This achieves 4.96\% and 2.94\% mean relative errors for $\Omega_m$ and $\sigma_8$ respectively. We did not spend much time on hyperparameter optimization for this network, so it is possible that marginally better results could be achieved. We achieve 5.24\% and 4.03\% using the 8 channel version of our model. We highlight that the information bottleneck here is just 8 floating point numbers, as compared to the 65,536 pixels in the original field data. The full output of the encoder (\textit{compressed field} in Fig.~\ref{fig:architecture}) is not used for parameter inference, only the 8 element summary statistics. The 16-channel model's latents achieve 3.72\% and 3.00\% mean relative errors, but exhibits worse frequency disentanglement.
We plot the results of parameter inference as we vary the number of latent channels in Fig.~\ref{fig:param-est-mse}.

% \vspace{-0.6em}

\begin{figure}[h!]
    \centering
    % \vspace{-1em}
    \includegraphics[width=0.5\linewidth]{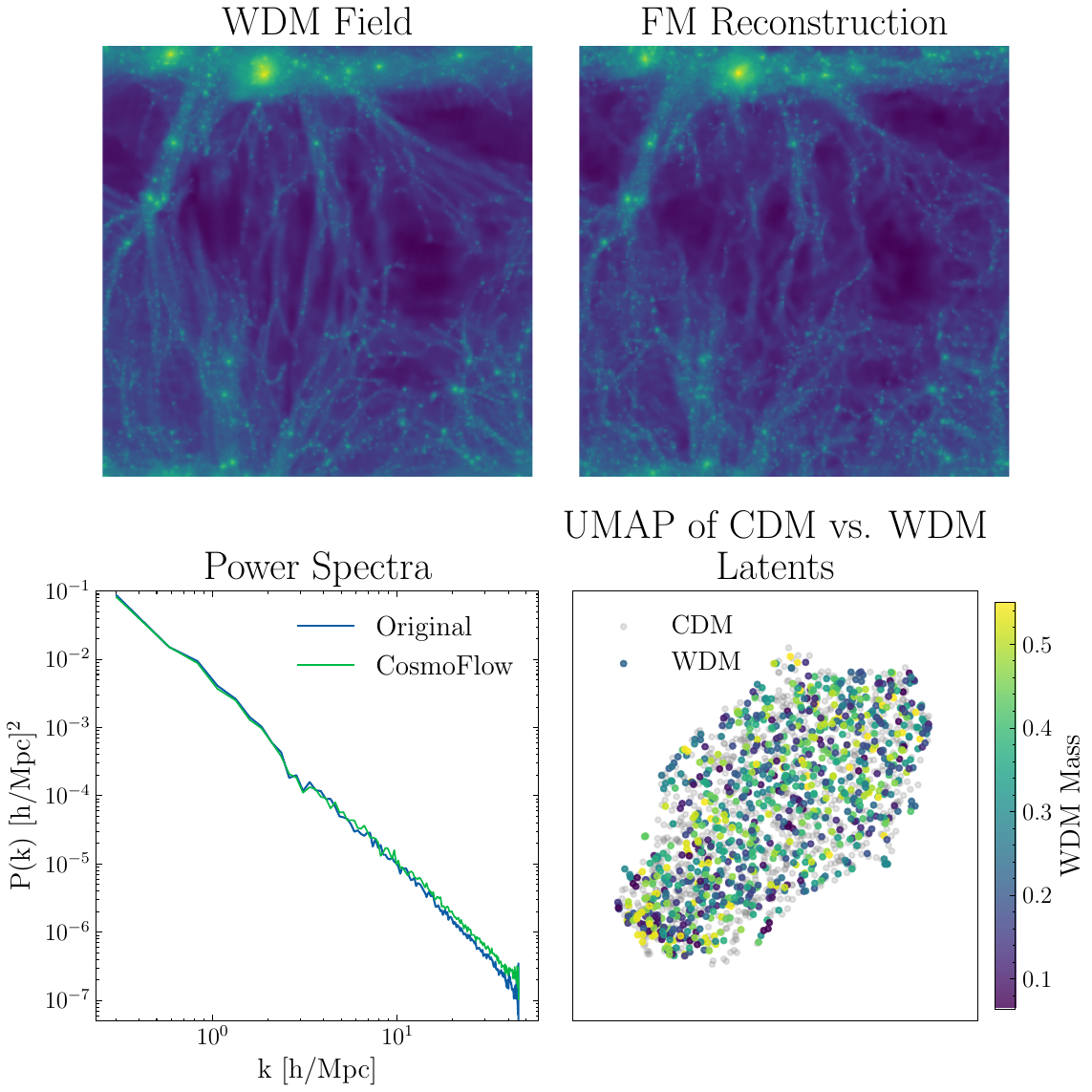}
    % \vspace{-0.7em}
    \caption{Warm dark matter maps are converted to ``nearest" cold dark matter maps. In particular, we observe \OURS artificially adds in fine structure, and the power spectrum over-represents high frequencies. The latent representations of CDM and WDM are not separable by UMAP.}
    \label{fig:wdm-to-cdm}
    \vspace{-1em}
\end{figure}

\paragraph{Anomaly Detection}
Another task of interest is \textit{anomaly detection}. In particular, we attempt to distinguish between cold dark matter and warm dark matter (WDM) maps using summary statistics. We observe that a) the model takes input WDM images, and then converts them to be in-distribution, CDM maps, as illustrated in Fig.~\ref{fig:wdm-to-cdm} and b) the latent representations of WDM maps are not easily separable from the latent representations of CDM maps.

\paragraph{Frequency-Based Interpolation/Latent Space Disentanglement}

Our model is designed to provide a disentangled representation, where different channels in the compressed field correspond to different spatial scales in the reconstruction. This gives us the ability to modulate the large scale structures \textit{independently} of the fine details, simply by interpolating the latent channels corresponding to those frequencies (See Fig. \ref{fig:interpolation}). As we increase the number of latent channels, this separability degrades; this is likely due to the latent space having ``too much" capacity, leading to latent channels encoding somewhat redundant information.

\section{Discussion and Future Work}
In this work, we demonstrate the utility of flow matching for learning representations of CDM simulation data. However, our current representations do not effectively capture features that distinguish CDM from WDM fields, and improving this remains a key goal, particularly to enable anomaly detection. We also plan to evaluate our representations in transfer learning settings, investigating whether they can be fine-tuned for new datasets with limited samples. To support compression and reconstruction, we aim to incorporate neural compression modules~[\cite{balle2018variational, yang2023lossy}], which may help reduce data size further. Finally, we see potential in extending the flow matching approach to other data modalities, such as directly operating on raw point cloud data instead of field-level maps.

\section{Software}
The code used to produce these results is accessible at \hyperlink{https://github.com/sidk2/cosmo-compression}{https://github.com/sidk2/cosmo-compression}.
% \FloatBarrier

%%
%% The next two lines define the bibliography style to be used, and
%% the bibliography file.
\bibliographystyle{ACM-Reference-Format}
\bibliography{references}

%%
%% If your work has an appendix, this is the place to put it.
% \newpage
\appendix
\newpage
\section{Architecture and Training Details}
In this section, we provide more details on the architecture and training hyperparameters.
\subsection{CosmoFlow Model}
\subsubsection{Encoder}
The encoder is a ResNet. Each ResNet block is comprised of [3x3 Conv with circular padding, BatchNorm, 3x3 Conv with circular padding, BatchNorm]. There is a residual connection from the input to the output. The encoder has 6 such ResNet blocks, along with an input convolutional layer (3x3 Conv, BatchNorm, MaxPool), and an output convolution layer (3x3 Conv). The input convolution produces 64 channels. The intermediate ResNet Blocks result in [64, 64, 128, 128, 256, 256] channels, respectively. The output convolution reduces this to 8 channels. This produces the compressed field. We then pass the compressed field through an adaptive average pooling layer to produce the summary statistics.

\subsubsection{Decoder}

The decoder is a UNet, with 4 downsampling stages and 4 upsampling stages. Each stage consists of four convolutional layers interspersed with batch normalization layers and GeLU activations.

We use sinusoidal positional encoding for the timestep embedding. The summary statistics are passed through a linear layer. These two vectors are used as conditioning for each downsampling and upsampling stage; more precisely, these are passed through an adaptive group normalization operation, then used for channel-wise modulation of the convolutional layer output, as described in [\cite{hudson2023sodabottleneckdiffusionmodels}]. Self attention modules are used after the second and third downsampling layer, as well as the second upsampling layer.

\subsubsection{Training Details}
The model is trained for 150 epochs. We use 14,000 samples from the Astrid set for training. The model is trained with the AdamW optimizer [\cite{loshchilov2019decoupledweightdecayregularization}], with the learning rate $\gamma =0.00005$, $\lambda=0.01$. We schedule the learning rate to decrease by a factor of 2 whenever the loss plateaus for 10 epochs.

\subsection{VAE Model}

We compare \OURS to a VAE model with same number of latent dimension. We adopt the encoder and decoder of a VAE image compression model proposed in~[\citep{balle2016end}]. The encoder is a sequence of 4 downsampling convolution layers with 5x5 kernels, each followed by generalized divisive normalization (GDN) layer~[\cite{balle2016end}] except for the last one. The first three convolution layers each has 128 channels, and the last one has 8 channels, resulting in a encoder output of size $8\times16\times16$. Two linear layers each with 1024 outputs are used to estimate the mean $\mu$ and log-variance $\log(\sigma^2)$ of the 1024 latent variables. The decoder mirrors the encoder in architecture, and is composed of 4 upsampling convolutions each followed by an inverse GDN layer. 

The model is trained with batch size of 256 for 300 epochs, using AdamW optimizer with reduced learning rate on plateau. 

\subsection{Parameter Inference}
\subsubsection{Parameter Inference on the Summary Statistics}
For parameter inference on the summary statistics, we use a fully connected network. We employ Optuna [\cite{optuna_2019}] for hyperparameter optimization. We independently optimize the network hyperparameters for each summary statistics size. For the 8-channel model, we use a single layer network with 2039 neurons. This is trained with the AdamW optimizer, with $\gamma = 1.85\times10^{-4}, \lambda=1.09\times10^{-7}$, for 200 epochs. However, we observe that the choice of hyperparameters has little effect on the results.

\subsubsection{Parameter Inference on the Raw Fields}
To do parameter inference on the raw fields, we use a modified version of the ResNet-18 architecture from [\cite{he2016deep}]. The output of the ResNet is modified to be a 256-dimensional vector, and then a fully-connected layer is added to project it down to 2 dimensions for the output. The model is trained with the AdamW optimizer, with the following parameters: $\gamma=0.0002$, $\beta=(0.5, 0.999)$, $\lambda=0.01$. It is trained for 200 epochs, and we use cosine annealing with $\gamma_{min} = 2\times10^{-6}$.
\section{Flow Matching Reconstruction Process}\label{sec:reconstruction-process}
\begin{figure*}[h]
    \includegraphics[width=\textwidth]{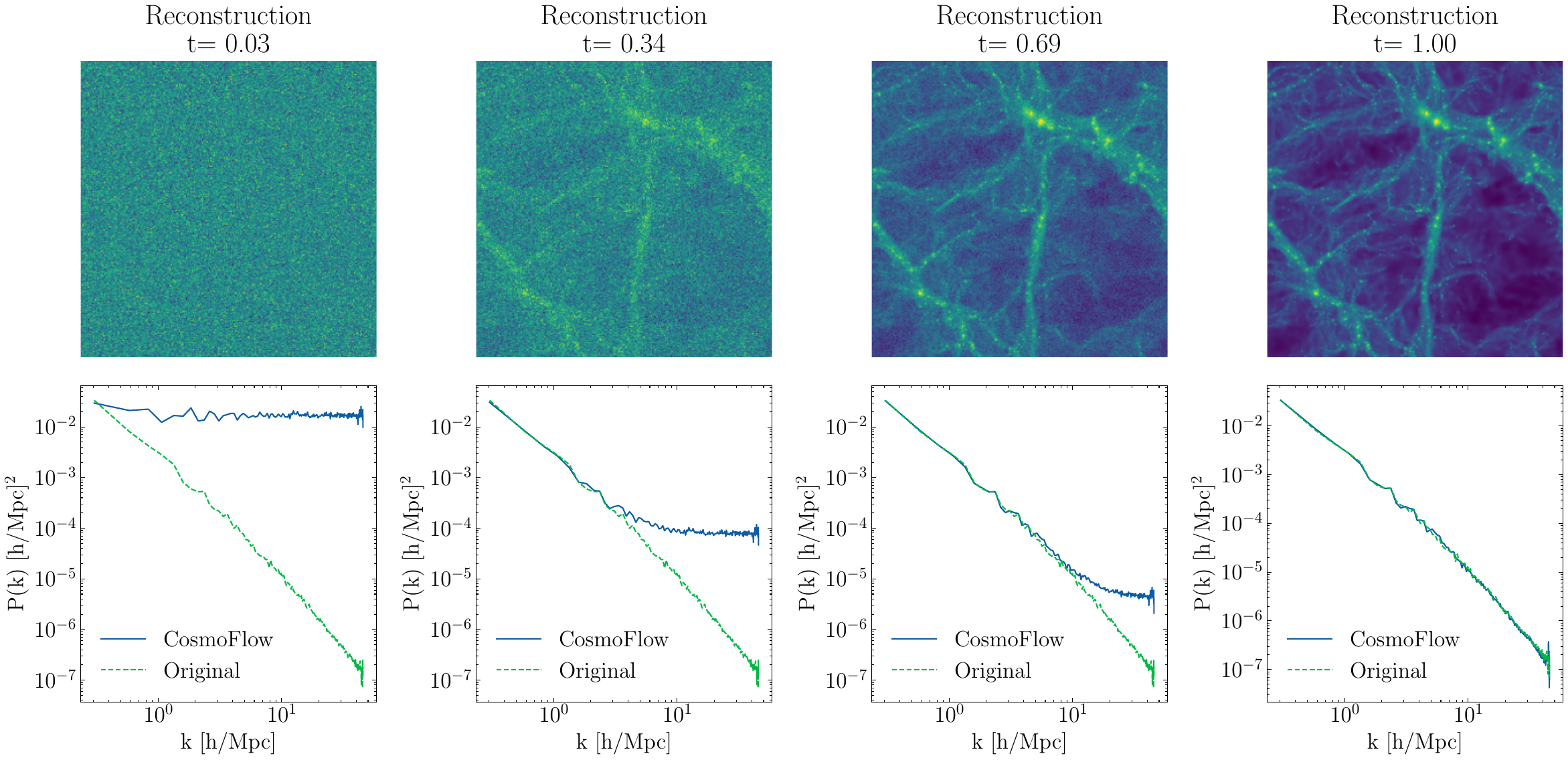}
    \caption{The generation process of a flow matching model. The model starts with a noise sample, and iteratively adds in structure. The power spectrum reflects this, starting as a flat spectrum, before adding in low frequencies and then the high frequencies. Note that for visualization purposes, we normalize the reconstructed power spectrum to match the amplitude of the target at low frequencies at all time steps. This is done by dividing the reconstructed power spectrum by $t^2$.}
    \label{fig:reconstruction_trajectory}
\end{figure*}

% Here, we show some further figures and results. 
In Fig.~\ref{fig:reconstruction_trajectory}, we demonstrate the inductive bias of flow matching. Images start out as Gaussian noise with flat power spectra. Large-scale features are constructed first, before the small-scale features are added in.

\section{Interpolation of CDM Fields}
\begin{figure*}[ht!]
    \includegraphics[width=\textwidth]{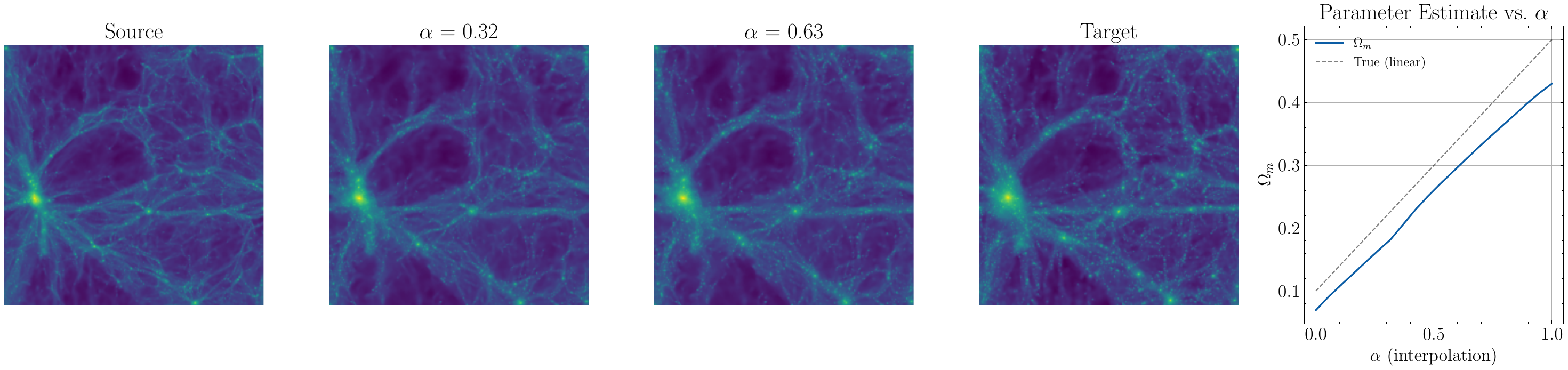}
    \caption{Interpolation between $\Omega_m = 0.1$ to $\Omega_m = 0.5$. Samples remain realistic throughout, and smoothly vary. The parameter inference network also shows a smooth increase in estimated value of $\Omega_m$. Parameter inference was done on the latent representation.}
    \label{fig:linear_interp}
\end{figure*}

In Fig.~\ref{fig:linear_interp}, we show the results of linearly interpolating between two CDM fields from the 1P dataset. These samples have exactly the same initial conditions and cosmological parameters, with the exception of the value of $\Omega_m$. We demonstrate that by linearly interpolating the latent space, we can generate realistic samples at intermediate values of $\Omega_m$, which we cross-validate by showing that our parameter inference network predicts a continuously increasing value for $\Omega_m$ over the course of the interpolation.
\end{document}